\documentclass{aa}

\newcommand{\nustar}{\textit{NuSTAR}}
\newcommand{\xrism}{\textit{XRISM}}

\newcommand{\src}{GS~1354--64}

\usepackage{amsmath}
\usepackage{graphicx}
\usepackage{parskip}
\usepackage{float}
\usepackage{graphicx}
\usepackage{multirow}
\usepackage{txfonts}
\usepackage{xspace}
\usepackage{xcolor}
\usepackage{url}
\usepackage[colorlinks=true, allcolors=blue]{hyperref}
\usepackage[flushleft]{threeparttable}
\usepackage{orcidlink}
\usepackage{seqsplit}

\begin{document}

\title{A clean broad iron line in GS 1354--64 as seen by \textit{XRISM}}

\author{Honghui Liu \inst{\ref{in:Tub}}\orcidlink{0000-0003-2845-1009}
\and Lingda Kong \inst{\ref{in:nankai}}\orcidlink{0000-0003-3188-9079}
\and Oluwashina K. Adegoke \inst{\ref{in:caltech}}\orcidlink{0000-0002-5966-4210}
\and Jiachen Jiang \inst{\ref{in:warwick}}\orcidlink{0000-0002-9639-4352}
\and Cosimo Bambi \inst{\ref{in:fudan},\ref{in:tash}}\orcidlink{0000-0002-3180-9502}
\and Andrew C. Fabian \inst{\ref{in:cam}}\orcidlink{0000-0002-9378-4072}
\and Adam Ingram \inst{\ref{in:Newcastle}}\orcidlink{0000-0002-5311-9078}
\and Swati Ravi \inst{\ref{in:mit}}\orcidlink{0000-0002-2381-4184}
\and James F. Steiner \inst{\ref{in:cfa}}\orcidlink{0000-0002-5872-6061}
\and Qingcang Shui \inst{\ref{in:ihep}}\orcidlink{0000-0001-5160-3344}
\and Dominic J. Walton \inst{\ref{in:Hertfordshire}}\orcidlink{0000-0001-5819-3552}
\and Yerong Xu \inst{\ref{in:ice}, \ref{in:ieec}}\orcidlink{0000-0002-2523-5485}
\and Andrew J. Young \inst{\ref{in:bristol}} \orcidlink{0000-0003-3626-9151}
\and Yuexin Zhang \inst{\ref{in:cfa}}\orcidlink{0000-0002-2268-9318}
\and Zuobin Zhang \inst{\ref{in:oxford}}\orcidlink{0000-0003-0847-1299}
\and Andrea Santangelo \inst{\ref{in:Tub},\ref{in:fudan}}\orcidlink{0000-0003-4187-9560}
}

\institute{
Institut f\"ur Astronomie und Astrophysik, Universit\"at T\"ubingen, Sand 1, D-72076 T\"ubingen, Germany \label{in:Tub} \\ \email{honghui.liu@uni-tuebingen.de}
\and School of Physics, Nankai University, Tianjin 300071, China \label{in:nankai}
\and Cahill Center for Astronomy \& Astrophysics, California Institute of Technology, Pasadena, CA 91125, USA \label{in:caltech}
\and Department of Physics, University of Warwick, Gibbet Hill Road, Coventry CV4 7AL, UK \label{in:warwick}
\and Center for Astronomy and Astrophysics, Department of Physics, Fudan University, Shanghai 200438, China \label{in:fudan}
\and School of Humanities and Natural Sciences, New Uzbekistan University, Tashkent 100001, Uzbekistan \label{in:tash}
\and Institute of Astronomy, University of Cambridge, Madingley Road, Cambridge CB3 0HA, UK \label{in:cam}
\and School of Mathematics, Statistics, and Physics, Newcastle University, Newcastle upon Tyne NE1 7RU, UK \label{in:Newcastle}
\and MIT Kavli Institute for Astrophysics and Space Research, Massachusetts Institute of Technology, 77 Massachusetts Avenue, Cambridge, MA 02139, USA \label{in:mit}
\and Center for Astrophysics \textbar\ Harvard \& Smithsonian, 60 Garden Street, Cambridge, MA 02138, USA \label{in:cfa}
\and Key Laboratory of Particle Astrophysics, Institute of High Energy Physics, Chinese Academy of Sciences, 100049, Beijing, People's Republic of China \label{in:ihep}
\and Centre for Astrophysics Research, University of Hertfordshire, College Lane, Hatfield AL10 9AB, UK \label{in:Hertfordshire}
\and Institute of Space Sciences (ICE, CSIC), Campus UAB, Carrer de Magrans, 08193 Barcelona, Spain \label{in:ice}
\and Institut d'Estudis Espacials de Catalunya (IEEC), Edifici RDIT, Campus UPC, 08860 Castelldefels, (Barcelona), Spain \label{in:ieec}
\and H.H. Wills Physics Laboratory, Tyndall Avenue, Bristol BS8 1TL, UK \label{in:bristol}
\and Astrophysics, Department of Physics, University of Oxford, Keble Road, Oxford OX1 3RH, UK \label{in:oxford}
}

\titlerunning{GS 1354--64 as seen by \textit{XRISM}}
\authorrunning{H.~Liu et al.}


\abstract{We present a spectroscopic analysis of \textit{XRISM} and \textit{NuSTAR} observations of the black hole X-ray binary GS~1354--64 during its 2026 outburst. A total number of 3.5 million photons are collected by the microcalorimeter Resolve on board \textit{XRISM}, providing an unprecedented high-resolution view of the iron line profile. A clean broad iron line is found in the data, without significant narrow features. Modeling the broad iron line with relativistic reflection from the inner accretion disk suggests a rapidly spinning black hole ($a_*>0.98$) in the system. Measurements of the disk inclination angle from the reflection method are model-dependent. This work demonstrates the power of X-ray microcalorimeters in studying the inner accretion flow and constraining black hole parameters.
}


\maketitle
\nolinenumbers

\section{Introduction} \label{sec:intro}

Accretion onto black holes is a fundamental process in astrophysics, powering some of the most energetic phenomena in the universe. In black hole X-ray binaries (BH XRBs), matter from a companion star is accreted onto a stellar-mass black hole, forming an accretion disk \citep{Shakura1973,Novikov1973blho.conf..343N}. The inner regions of the disk emit thermal radiation that peaks in the soft X-ray band, while a hot corona produces a non-thermal power-law component through inverse Compton scattering \citep{Shapiro1976ApJ...204..187S, Galeev1979ApJ...229..318G, Titarchuk1994ApJ...434..570T}. BH XRBs exhibit distinct spectral states, with the hard state dominated by coronal emission and the soft state dominated by disk emission \citep{Remillard2006, Done2007}.

In addition to the direct emission from the disk and corona, a fraction of the coronal photons is intercepted and reprocessed by the accretion disk, producing a reflection component. The local reflection spectrum is characterized by fluorescent emission lines (the most prominent being the iron K$\alpha$ line at 6.4~keV), absorption edges, and a Compton hump peaking around 30~keV \citep{George1991, Garcia2010}. These emission lines are smeared and broadened \citep{Fabian2000, Dauser2010} by the orbital motion of the disk material and gravitational redshift near the black hole \citep{Fabian1989, Gates2025PhRvD.111l4004G}. The observed reflection spectrum therefore exhibits a relativistically-broadened iron line and a Compton hump. The detailed profile of this line depends on several factors, including the black hole spin \citep{Brenneman2006ApJ...652.1028B, Bambi2021}, disk inclination angle \citep{Garcia2014, Liu2025MNRAS.536.2594L, Huang2025ApJ...989..168H}, coronal geometry \citep{Wilkins2011MNRAS.414.1269W, Dauser2013}, and the physical properties of the disk such as density \citep{Garcia2016,Jiang2019gx339,Liu2023ApJ...951..145L}, ionization state \citep{Garcia2010}, and elemental abundances. Modeling the relativistic reflection spectrum enables measurements of these key parameters \citep{Brenneman2006ApJ...652.1028B, Garcia2015, Jiang2019gx339, Liu2022, Liu2023ApJ...950....5L}.

Previously, X-ray reflection spectroscopy has mainly been performed with CCD detectors, which have a spectral resolution of around 150 eV at 6 keV \citep[e.g.,][]{Fabian2012MNRAS.424..217F,Walton2013MNRAS.428.2901W,Walton2016}. The limited spectral resolution makes it difficult to resolve narrow emission and absorption features on top of the broad iron line. This might be the reason behind ongoing debates over modeling the reflection component in BH XRBs. For example, in some systems it remains unclear whether a narrow component should be included when fitting the spectra, which may arise from reflection by distant material \citep[e.g.,][]{Garcia2015, Liu2023ApJ...950....5L}. In some cases, a dual corona configuration has been suggested to fit the data \citep[e.g.,][]{Zdziarski2022ApJ...928...11Z}. These scenarios that introduce subtle features to the reflection spectrum can be better tested with high energy-resolution data \citep[e.g.,
][]{Liu2025MNRAS.536.2594L}.

The X-ray Imaging and Spectroscopy Mission (\xrism{}; \citealt{Tashiro2021SPIE11444E..22T}) is equipped with the Resolve instrument, a high-resolution X-ray microcalorimeter that provides an energy resolution of about 5 eV at 6 keV. \xrism{} has confirmed broadened iron line features in a BH XRB (Cygnus X-1, \citealt{Draghis2025ApJ...995L..12D}) and an active galactic nucleus (MCG-6-30-15, \citealt{Brenneman2025ApJ...995..200B}). In both cases, the broad iron line is superimposed with narrow features and the high-resolution data from Resolve could disentangle the broad and narrow features.

\src{} is a BH XRB system discovered in 1987 \citep{Kitamoto1990ApJ...361..590K}. It subsequently underwent outbursts in 1997 \citep{Revnivtsev2000ApJ...530..955R} and 2015 \citep{Miller2015ATel.7612....1M,El-Batal2016ApJ...826L..12E}. The 1997 and 2015 outbursts remained in the hard state only and are classified as a so-called ``failed outburst'' \citep{Stiele2016MNRAS.459.4038S}. Strong disk reflection features were observed during the 2015 outburst with \nustar{}, constraining the black hole spin to be high ($a_*>0.99$; \citealt{El-Batal2016ApJ...826L..12E, Xu2018ApJ...865..134X}). The disk inclination inferred from reflection modeling is model-dependent, ranging from $20^\circ$ to $70^\circ$ \citep{Liu2023ApJ...951..145L,Draghis2024ApJ...969...40D,El-Batal2016ApJ...826L..12E, Xu2018ApJ...865..134X}. Radial-velocity measurements yield an orbital period of about 2.5 days and a mass function of $5.7 \pm 0.3\,M_{\odot}$ \citep{Casares2009ApJS..181..238C}. The absence of eclipses constrains the binary inclination to be below $79^\circ$ \citep{Casares2009ApJS..181..238C}. While a lower limit of 25~kpc has been proposed for the distance \citep{Casares2009ApJS..181..238C}, parallax measurements by Gaia suggest a much closer value of $0.7_{-0.3}^{+3.4}$~kpc \citep{Gandhi2019MNRAS.485.2642G}.

In this paper, we present a new spectroscopic analysis of \xrism{} and \nustar{} \citep{Harrison2013ApJ...770..103H} observations of \src{} during its 2026 outburst. This new outburst reached a peak flux six times higher than the 2015 outburst in the MAXI band and shows indication of transition to the soft state (see Figure~\ref{HID} and \citealt{2026ATel17638....1S}). \xrism{} and \nustar{} observed the source during the bright hard state, at a flux level twice the peak flux of the 2015 outburst. The high energy-resolution Resolve instrument could provide a detailed view of the iron band and the \nustar{} data could constrain the broadband continuum and the Compton hump.

The paper is organized as follows. In Section~\ref{data} we describe the data reduction process. The spectral fitting results are presented in Section~\ref{analysis}. We discuss the results in Section~\ref{discuss}.

\section{Observations and Data Reduction}
\label{data}

\subsection{\xrism{}}

\xrism{} observed \src{} from 20 to 23 January 2026 \citep{Liu2026ATel17625....1L} with both the Resolve \citep{2022SPIE12181E..1SI} and Xtend  \citep{2022SPIE12181E..1TM} detectors. The exposure time is around 130 ks for Resolve. The observation is accompanied by a \nustar{} observation (see Table~\ref{tab:obs} for details). Figure~\ref{HID} shows the hardness-intensity diagram of \src{} from MAXI data, with the time of the \xrism{} observation marked.

The data are reduced following the standard data reduction procedure as outlined in the \xrism{} Data Analysis Guide\footnote{\url{https://heasarc.gsfc.nasa.gov/docs/xrism/analysis/}}. We use the tools in HEASOFT v6.36 and the calibration database (CALDB) version 20250915. Spectra and light curves are extracted using the \texttt{xselect} tool. 

For Resolve, only high-resolution primary events (GRADE=0) are selected, and events from pixels 12 and 27 are excluded. We use the \texttt{rslmkrmf} tool to generate the extra-large (X) response matrix. The \texttt{xaexpmap} task is then used to make the exposure map. The Auxiliary Response File (ARF) is generated with the \texttt{xaarfgen} tool.

For Xtend, the observation is performed in the 1/8 window + 0.1 s burst mode. The data are affected by pile-up. We extract the source spectrum from an annular region with inner and outer radii of 25 and 35 pixels, respectively. The inner radius is chosen such that increasing it further does not change the spectral shape. In this case, the pile-up fraction should be lower than 5\%\footnote{Fig 7.1 of the XRISM ABC guide shows the simulated pile-up fraction as a function of excluding radius: \url{https://heasarc.gsfc.nasa.gov/docs/xrism/analysis/abc_guide/Xtend_Data_Analysis.html}.}. The background spectrum is extracted from a source-free circular region with a radius of 35 pixels. The response matrix is generated using the \texttt{xtdrmf} tool and the ARF file is calculated using the \texttt{xaarfgen} tool. 

The spectra are grouped using \texttt{ftgrouppha}\footnote{\url{https://heasarc.gsfc.nasa.gov/lheasoft/help/ftgrouppha.html}} with the options ``\texttt{grouptype=optmin groupscale=5}''. This applies the ``optimal binning'' strategy \citep{Kaastra2016A&A...587A.151K} while also ensuring a minimum of 5 counts per bin. The data are fitted in the 2--12.5 keV energy range for Resolve and 3--10 keV for Xtend. Data below 3 keV are not considered for Xtend because discrepancies have been identified in this energy range between Xtend and XMM-Newton/PN \citep[e.g.,][]{Brenneman2025ApJ...995..200B, 2025A&A...702A.147X}. This issue is still under investigation. We also find a similar discrepancy between Resolve and Xtend in the 2--3 keV range.

\begin{table*}
\centering
\renewcommand{\arraystretch}{1.8}  

\begin{tabular}{ccccccc}
\hline
Mission  & Obs start & Obs end & Observation ID & Exposure (ks) & Counts (10$^6$) \\ 
\hline \hline
\xrism{}    & 2026-01-20T04:47:07 & 2026-01-23T03:26:44 & 902002010 & 128.6 & 3.48 \\
\nustar{}   & 2026-01-21T07:06:09 & 2026-01-21T19:01:09 & 91201302004 & 25.8 & 3.37\\
\hline\hline 
\end{tabular}

\caption{Log of the \xrism{} and \nustar{} observations of \src{} analyzed in this work. The exposure time and counts are given for the Resolve instrument of \xrism{} and both modules of \nustar{}.}
\label{tab:obs}
\end{table*}

\subsection{\nustar{}}

We reduce the \nustar{} mode 1 data using \texttt{NUSTARDAS} from \texttt{HEASOFT} v6.36. Cleaned event files for both FPMA and FPMB are produced with the \texttt{nupipeline} tool and the calibration database (CALDB) version 20241104. The bright-source flag is implemented by setting \texttt{statusexpr="(STATUS==b0000xxx00xxxx000) \&\& (SHIELD==0)"}. We extract the source spectra from a circular region with a radius of 180 arcsec and the background spectra from a source-free polygonal region. The spectra are grouped using the same method as described above. The data are fitted in the 4--79 keV energy range\footnote{Data in the 3--4 keV range from this observation are not consistent between FPMA and FPMB.}.

\subsection{Strictly simultaneously datasets}

Figure~\ref{lc} presents the lightcurves of \xrism{}/Resolve and \nustar{}/FPMA with a time resolution of 100~s. 
We extract spectra from time intervals that are strictly simultaneous between \xrism{} and \nustar{} (e.g., the red data in Figure~\ref{lc}). The simultaneous intervals are determined by cross-matching the good time intervals (GTIs) of Resolve and FPMA. The spectra are extracted from the same regions described above, using only events that fall within these simultaneous intervals. New response files are generated. In this simultaneous dataset, there are $2.4\times10^5$ counts for Resolve and $1\times10^6$ counts for FPMA and FPMB combined. The spectra are then grouped using the same strategy described above for spectral fitting.

\begin{figure}[htbp]
    \centering
    \includegraphics[width=0.49\textwidth]{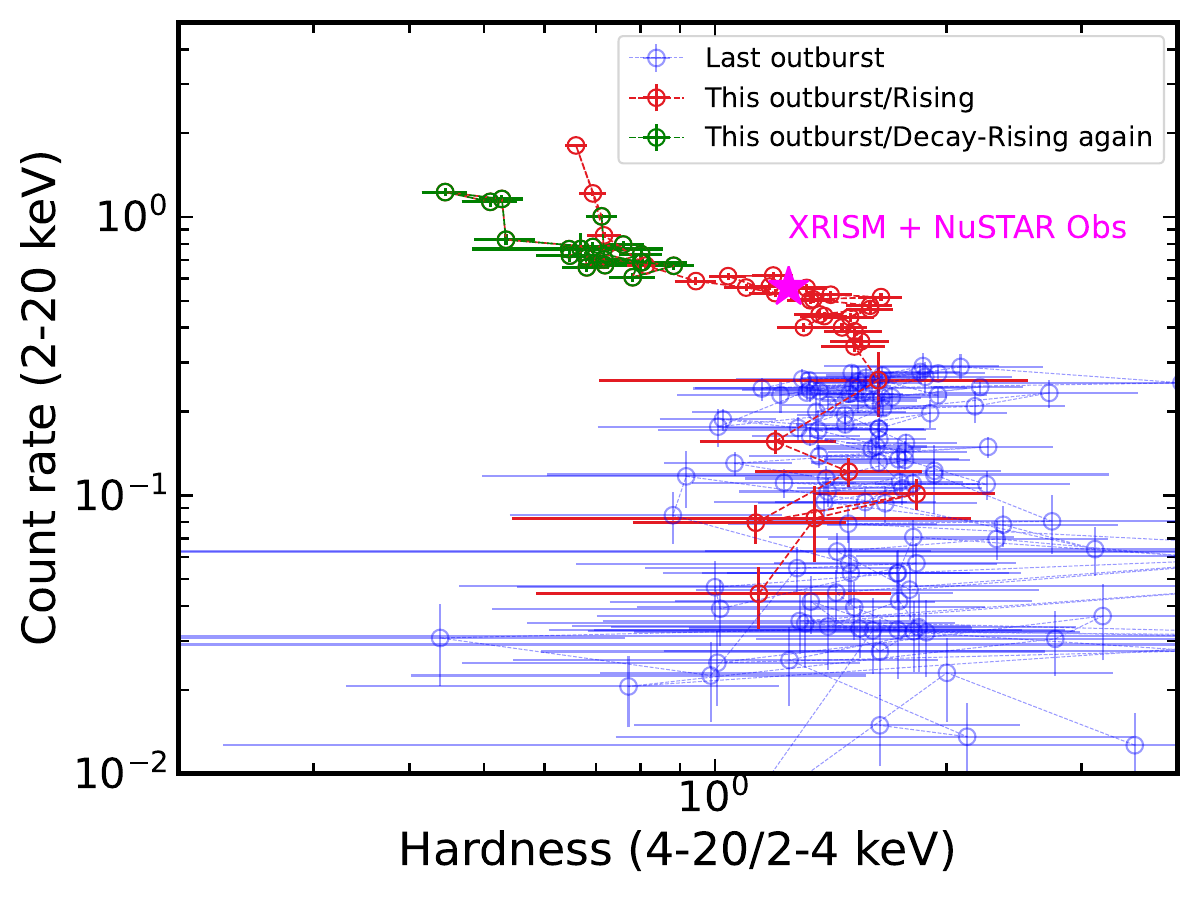}
    \caption{
        The hardness-intensity diagram for \src{} from MAXI data. The blue data represent the 2015 outburst, while the red and green data represent the 2026 outburst. The magenta star indicate the time of the \xrism{} and \nustar{} observations analyzed in this work.
    }
    \label{HID}
\end{figure}



\begin{figure*}[bhtp]
    \centering
    \includegraphics[width=0.8\textwidth]{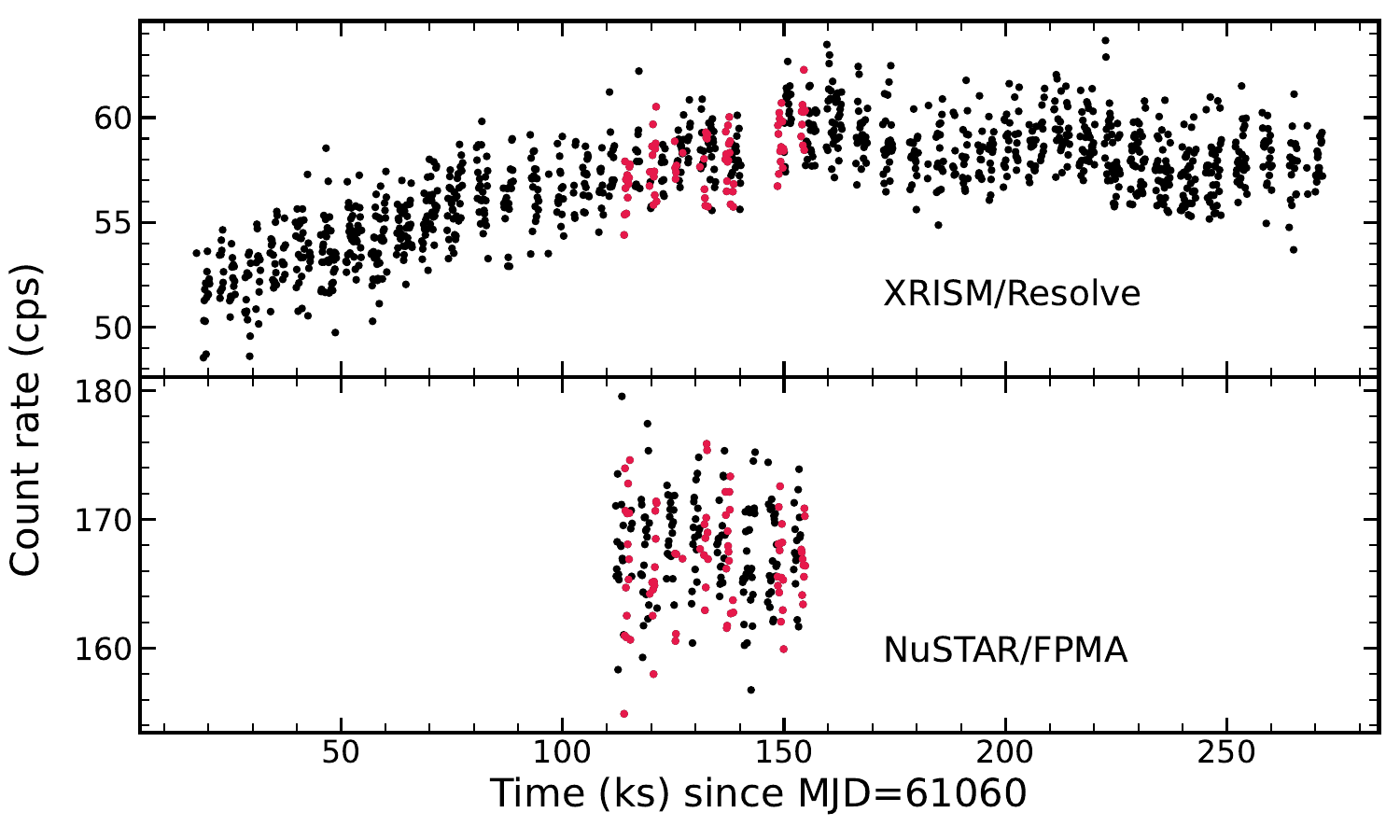}
    \caption{
        Lightcurve of the \xrism{} and \nustar{} observations of \src{}. The time resolution is 100~s. The red data represent overlapping time intervals between the two instruments. The simultaneous spectral dataset is extracted from these overlapping intervals.
    }
    \label{lc}
\end{figure*}


\section{Spectral Analysis}
\label{analysis}

We perform spectral analysis using \texttt{XSPEC} v12.15.1 \citep{xspec}. Throughout the analysis, we use elemental abundances from \cite{Wilms2000} and photoelectric cross-sections from \cite{Verner1996}. The Cash statistics \citep{Cash1979ApJ...228..939C, Kaastra2017A&A...605A..51K} are used to determine best-fit values and uncertainties (reported at the 90\% confidence level unless otherwise stated).

\subsection{Reflection features}

As a first step, we simultaneously fit the \xrism{}/Resolve, \xrism{}/Xtend, \nustar{}/FPMA and \nustar{}/FPMB spectra from the total dataset using a simple absorbed continuum model: \texttt{Constant*Tbabs*Cutoffpl}. The \texttt{Tbabs} \citep{Wilms2000} component takes the Galactic absorption into account, while the \texttt{Cutoffpl} component models the coronal emission with a power-law and an exponential high-energy cutoff. Since we are not considering any data below 2 keV, the column density of \texttt{Tbabs} is fixed at $7\times 10^{21}$~cm$^{-2}$ \citep{El-Batal2016ApJ...826L..12E}.

\begin{figure*}[htbp]
    \centering
    \includegraphics[width=0.95\textwidth]{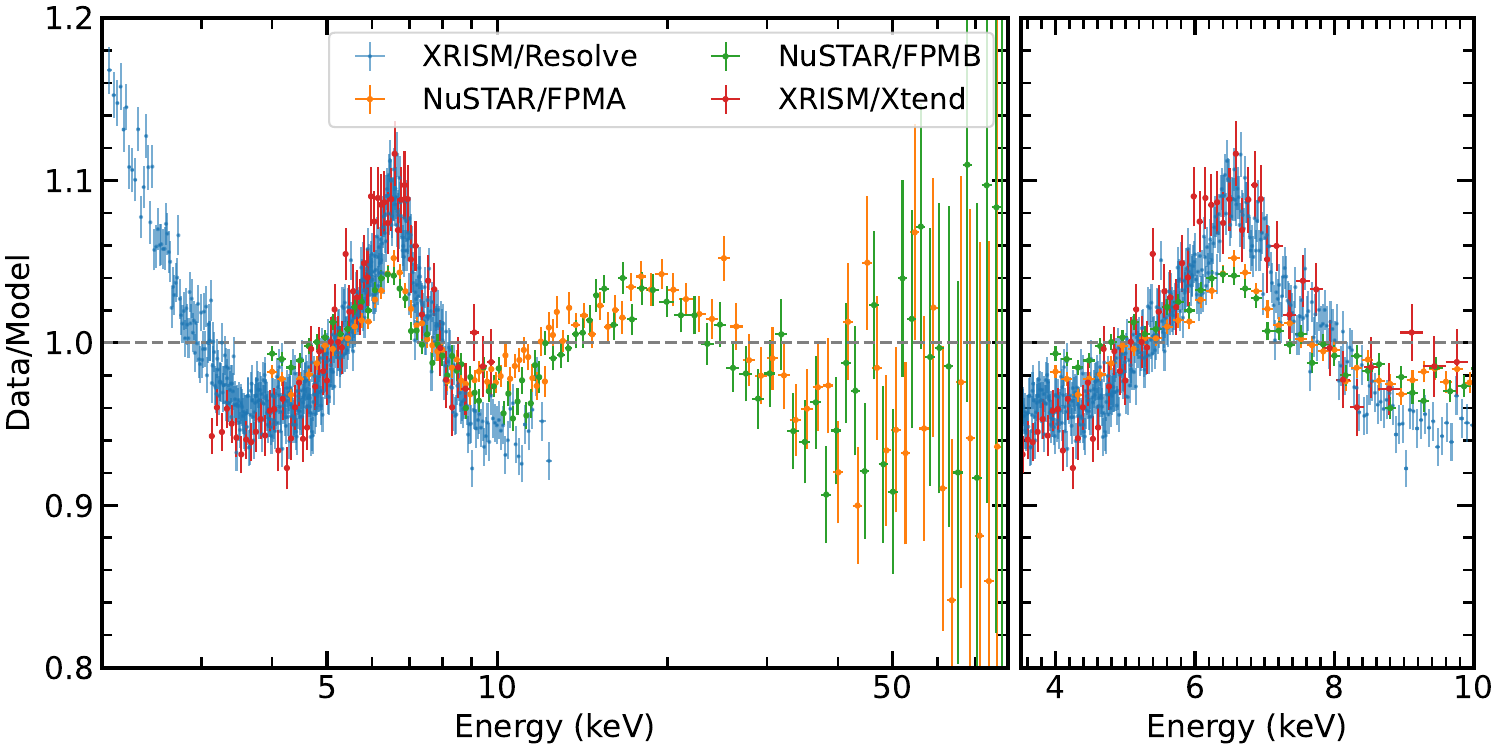}
    \caption{
        Reflection features in the \xrism{} and \nustar{} observations of \src{}. The data are fitted with a simple absorbed continuum model: \texttt{Tbabs*Cutoffpl}. The plot is only for illustration purposes and the data are binned for visual clarity. The broad excess around 6--7 keV and the hump near 30 keV indicate the presence of a relativistic reflection component.
    }
    \label{ironline}
\end{figure*}

The ratios of the data to the best-fit models are shown in Figure~\ref{ironline}. A broad excess peaking around 6--7 keV and a hump near 20 keV are evident. These features correspond to the broad iron line and the Compton hump, indicating the presence of a relativistic reflection component. Moreover, even with the energy resolution of Resolve, the iron line profile appears broad and singly-peaked, i.e. there are no additional narrow emission components, nor any apparent absorption lines from ionized iron introducing complex sub-structure.


\subsection{Reflection analysis}

\subsubsection{Simultaneous dataset}

To model the reflection component, we test two different models with different assumptions on the coronal geometry. We start with the simultaneous dataset (red data in Figure~\ref{lc}) to determine the baseline model. The models we test are as follows:

\begin{itemize}
    \item Model 1 (M1): \texttt{\seqsplit{Constant * TBabs * relxillCp}}
    \item Model 2 (M2): \texttt{\seqsplit{Constant * TBabs * relxilllpCp}}
\end{itemize}

The \texttt{relxillCp} model is a flavor of the disk reflection model package \texttt{relxill}\footnote{\url{https://www.sternwarte.uni-erlangen.de/~dauser/research/relxill/}} v2.8 \citep{Garcia2010, Garcia2013, Garcia2014}. It uses \texttt{nthComp} \citep{Zdziarski1996,Zycki1999} as the incident coronal spectrum irradiating the accretion disk. Other parameters in the model include: the black hole spin ($a_*$), the disk inclination angle ($i$), the emissivity profile, the iron abundance ($A_{\rm Fe}$), the inner disk radius ($R_{\rm in}$), the outer disk radius ($R_{\rm out}$), the reflection fraction ($R_{\rm f}$, \citealt{Dauser2016}), the disk ionization parameter ($\log\xi$), the disk density ($\log n_{\rm e}$) and the normalization. We fit the emissivity profile with a power-law. 

In Model 2, the \texttt{relxilllpCp} model assumes a lamppost geometry, where the corona is modeled as a point source located above the BH at a certain height ($h$). We set an upper limit of 20 $R_{\rm g}$ for the coronal height. The emissivity profile is calculated self-consistently from the lamppost geometry \citep{Dauser2013}. 
The reflection fraction is calculated self-consistently from the coronal geometry and the returning radiation effect is switched on \citep{Dauser2022}. Moreover, we switch on the ionization gradient in the disk, assuming the density gradient of an alpha disk (\texttt{iongrad\_type=2}).

For these initial fits, we fix the black hole spin at the maximum value of 0.998, which has been suggested by previous studies \citep{El-Batal2016ApJ...826L..12E, Xu2018ApJ...865..134X}. We also fix the outer disk radius at 400~$R_{\rm g}$ (where $R_{\rm g}=GM_{\rm BH}/c^2$ is the gravitational radius). This value of $R_{\rm out}$ is sufficiently large that increasing it further has no effect on the fit. The inner disk radius is left as a free parameter. We test whether a disk thermal component (\texttt{diskbb}; \citealt{Mitsuda1984}) is required in Model 1, but neither the inner disk temperature nor the normalization can be constrained by the data.

\begin{figure}[htbp]
    \centering
    \includegraphics[width=0.49\textwidth]{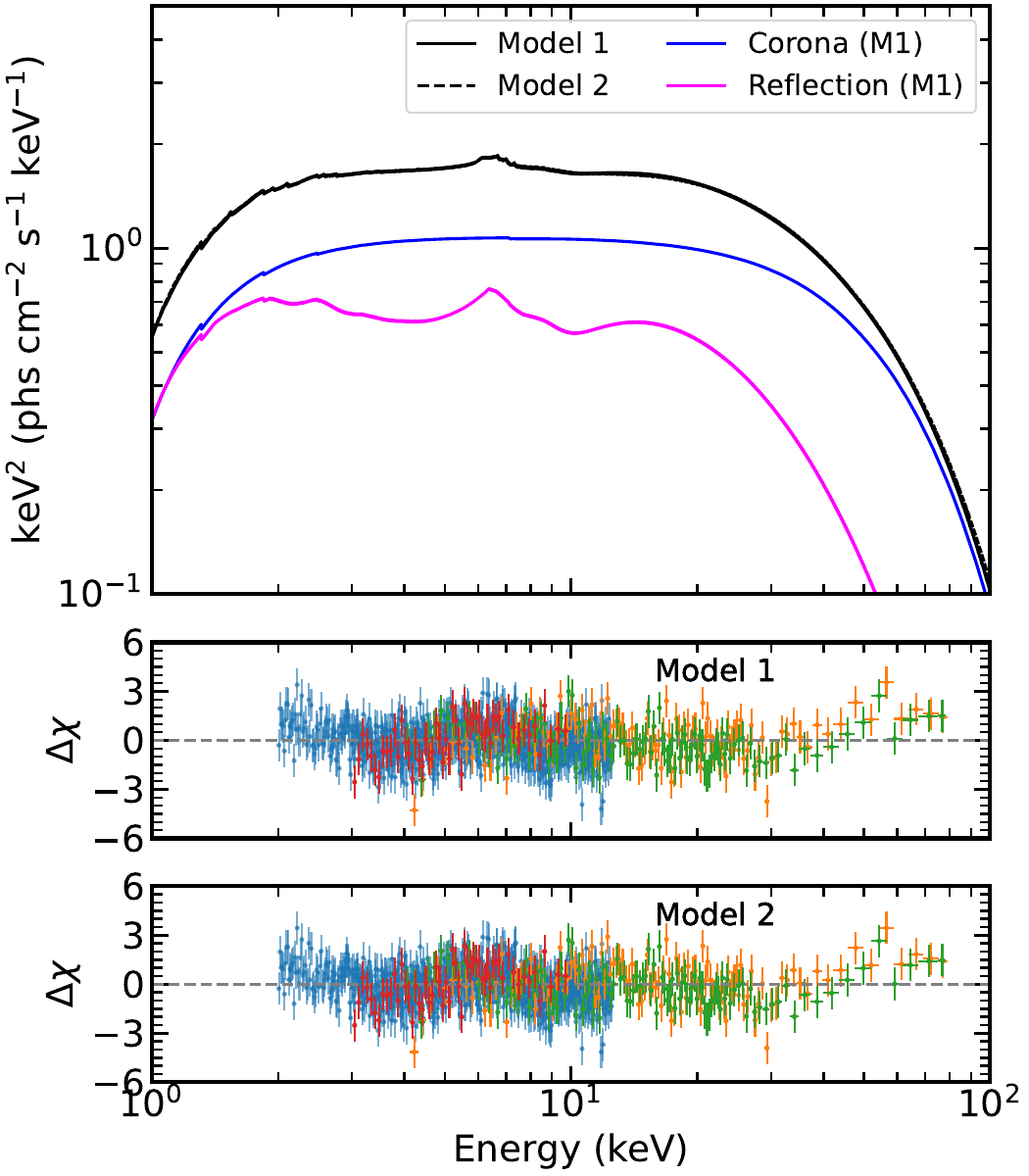}
    \caption{
        Total spectra (upper panel) and residuals (lower panels) for the strictly simultaneous dataset fitted with Models~1 and 2. In the upper panel, the separate coronal and reflection components for Model~1 are also shown. The data are binned for visual clarity. The color coding is the same as in Figure~\ref{ironline}.
    }
    \label{simul_resi}
\end{figure}

The best-fit parameters for the simultaneous dataset are presented in Table~\ref{tab:bestfit_combined} and the corresponding residuals are shown in Figure~\ref{simul_resi}. Both models provide good fits to the data, with Model 1 having the lowest C-stat value. The best-fit parameters from the models are broadly consistent with each other, except for the inclination angle. Model 1 suggests a higher inclination angle compared to Model 2. The iron abundance is sub-solar in both models. Assuming a maximally rotating BH, the disk appears to be slightly truncated, with the inner disk radius below $3~R_{\rm g}$. However, this can also be interpreted as a BH with $a_* \sim 0.8$, for which the disk extends down to the innermost stable circular orbit (ISCO). 

In the lamppost coronal model, the corona height is only loosely constrained ($h>13~ R_{\rm g}$). The coronal temperature is low ($\sim$~15~keV) and there is weak sign of unresolved high-energy tail in the \nustar{} data. This indicates the possibility of a hybrid corona with both thermal and non-thermal electrons \citep[e.g.,][]{Fabian2017}. Such a sign of hybrid corona has also been found in the 2015 outburst of \src{} \citep{Liu2023ApJ...951..145L}. 

\begin{table*}[htbp]
\centering
\renewcommand{\arraystretch}{1.8}
\setlength{\tabcolsep}{6pt}
\begin{tabular}{c cc |cccc}
\hline\hline
 & \multicolumn{2}{c}{Simultaneous dataset} & \multicolumn{3}{c}{Full dataset} \\
\cline{2-3}\cline{4-7}
Parameter & Model 1 & Model 2 & Model 1 & Resolve only & \multicolumn{2}{c}{NuSTAR only} \\
\hline

$\Gamma$
 & $1.997_{-0.021}^{+0.006}$
 & $1.962_{-0.008}^{+0.007}$
 & $1.776_{-0.007}^{+0.009}$
 & $1.810_{-0.011}^{+0.008}$ 
 & $1.760_{-0.026}^{+0.033}$
 & $1.851_{-0.005}^{+0.003}$\\

$kT_{\rm e}$ (keV)
 & $14.3_{-0.3}^{+0.3}$
 & $15.9_{-0.5}^{+0.3}$
 & $27_{-4}^{+2}$
 & $27^{*}$ 
 & $26.50_{-0.9}^{+14}$
 & --- \\

$E_{\rm cut}$ (keV)
 & ---
 & ---
 & ---
 & --- 
 & ---
 & $39.8_{-0.4}^{+0.8}$\\

$a_*$
 & $0.998^*$
 & $0.998^*$
 & $>0.99$
 & $>0.98$ 
 & $>0.98$ 
 & $0.910_{-0.07}^{+0.008}$ \\

Incl (deg)
 & $50_{-4}^{+3}$
 & $32_{-2}^{+2}$
 & $11_{-2}^{+4}$
 & $11_{-2}^{+2}$ 
 & $<15$ 
 & $59_{-5}^{+2}$ \\

$q_{\rm in}$
 & $7.5_{-1.2}^{+1.5}$
 & ---
 & $9.07_{-0.27}^{+0.5}$
 & $10.00_{-0.23}^{+P}$ 
 & $7.42_{-0.06}^{+1.1}$
 & $>7.8$\\

$q_{\rm out}$
 & $=q_{\rm in}$
 & ---
 & $2.61_{-0.06}^{+0.09}$
 & $2.54_{-0.08}^{+0.07}$ 
 & $2.79_{-0.15}^{+0.25}$
 & $=q_{\rm in}$\\

$R_{\rm br}$ ($R_{\rm g}$)
 & ---
 & ---
 & $3.57_{-0.22}^{+0.1}$
 & $3.39_{-0.07}^{+0.06}$ 
 & $4.47_{-0.19}^{+0.18}$ 
 & --- \\

$R_{\rm in}$ ($R_{\rm ISCO}$)
 & $2.11_{-0.11}^{+0.3}$
 & $2.3_{-0.5}^{+0.3}$
 & $<1.3$
 & $1.16_{-0.11}^{+0.28}$ 
 & $1^*$ 
 & $1^*$\\

$\log(\xi)$
 & $2.817_{-0.04}^{+0.021}$
 & $3.29_{-0.07}^{+0.7}$
 & $3.273_{-0.029}^{+0.05}$
 & $3.25_{-0.04}^{+0.04}$ 
 & $3.63_{-0.22}^{+0.05}$
 & $3.075_{-0.030}^{+0.016}$\\

$\log(n_{\rm e})$
 & $18.96_{-0.09}^{+0.05}$
 & $17.76_{-0.16}^{+0.18}$
 & $20.00_{-0.22}^{+P}$
 & $20.00_{-0.15}^{+P}$ 
 & $19.0_{-0.9}^{+0.3}$
 & $15^*$\\

$A_{\rm Fe}$ (solar)
 & $0.500_{-P}^{+0.014}$
 & $0.59_{-0.04}^{+0.04}$
 & $1.23_{-0.16}^{+0.21}$
 & $1.52_{-0.29}^{+0.4}$ 
 & $2.81_{-0.23}^{+1.2}$
 & $0.52_{-P}^{+0.08}$\\

$h$ ($R_{\rm g}$)
 & ---
 & $17.0_{-4}^{+P}$
 & ---
 & --- 
 & ---\\


$R_{\rm f}$
 & $1.05_{-0.12}^{+0.07}$
 & ---
 & $10.0_{-2.6}^{+P}$
 & $2.9_{-0.7}^{+1.1}$ 
 & $10_{-4}^{+P}$ 
 & $0.735_{-0.165}^{+0.087}$\\

$N_{\rm Refl}$
 & $0.1964_{-0.0008}^{+0.0011}$
 & $0.1943_{-0.0007}^{+0.0017}$
 & $0.1875_{-0.0004}^{+0.0004}$
 & $0.1876_{-0.0004}^{+0.0004}$ 
 & $0.2157_{-0.0015}^{+0.0033}$
 & $0.2135_{-0.0008}^{+0.0008}$\\

\hline
\nustar{} & \multicolumn{2}{c}{} & \multicolumn{2}{c}{} \\
$\log(\xi)$ & \multicolumn{2}{c}{} & $3.663_{-0.023}^{+0.024}$ & --- \\
$\log(n)$   & \multicolumn{2}{c}{} & $19.48_{-0.15}^{+0.09}$   & --- \\
\hline

$C_{\rm FPMA}$
 & $1.147_{-0.004}^{+0.004}$
 & $1.1431_{-0.005}^{+0.0025}$
 & $1.147^{*}$
 & --- 
 & $1^*$
 & $1^*$ \\

$C_{\rm FPMB}$
 & $1.150_{-0.005}^{+0.004}$
 & $1.1463_{-0.005}^{+0.0029}$
 & $1.150^{*}$
 & --- 
 & $1.0067_{-0.0017}^{+0.0017}$ 
 & $1.0067_{-0.0017}^{+0.0016}$\\

$C_{\rm Xtend}$
 & $1.16_{-0.01}^{+0.01}$
 & $1.16_{-0.01}^{+0.01}$
 & $1.158_{-0.004}^{+0.004}$
 & --- \\

\hline
Cstat/d.o.f
 & 3847.9/3628
 & 3867.3/3629
 & 4679.8/4408
 & 3964.7/3852 
 & 557.2/468
 & 561.1/471 \\
 
\hline\hline
\end{tabular}
\caption{Best-fit parameter values and their uncertainties for the exactly simultaneous dataset (Models~1--2) and the full dataset (Model~1; joint fit, Resolve-only fit and NuSTAR-only fit). For the simultaneous dataset, the black hole spin is fixed at 0.998 and the outer disk radius is fixed at 400~$R_{\rm g}$ for all models. The reflection fraction is calculated self-consistently for Models~2. Parameters marked with $^{*}$ are fixed.}
\label{tab:bestfit_combined}
\end{table*}

\subsubsection{Full dataset}

Since Model 1 provides the best fit to the simultaneous dataset, we apply it to the full dataset. The full dataset contains ten times more photon counts than the simultaneous dataset in Resolve, and we therefore expect tighter parameter constraints. We test the scenario in which both the black hole spin and the inner disk radius are free parameters. In the full dataset, the \xrism{} and \nustar{} data do not fully agree in the overlapping energy range. The main discrepancies lie in the iron band. This is likely due to the fact that the source is variable across the \xrism{} observation (see Figure~\ref{lc}). Therefore, we fix the cross-normalization factor between the \xrism{} and \nustar{} spectra at the best-fit value obtained from the simultaneous dataset. We then allow the ionization parameter and the disk density to differ between the \xrism{} and \nustar{} spectra. Alternative scenarios, such as unlinking the photon index, the normalization, and the reflection fraction parameters, have also been tested. These tests yield measurements of the BH spin, disk inclination angle, and inner disk radius that are consistent with those obtained when unlinking the ionization and density parameters.

We also fit the \xrism{}/Resolve-only and \nustar{}-only data with Model 1. In the former case, the Resolve data alone cannot constrain the coronal temperature, so we fix it at the best-fit value from the full dataset. The \nustar{}-only data cannot constrain both the black hole spin and the inner disk radius. Therefore, we fix $R_{\rm in}$ at the ISCO radius and fit only for the BH spin.

\begin{figure}[htbp]
    \centering
    \includegraphics[width=0.49\textwidth]{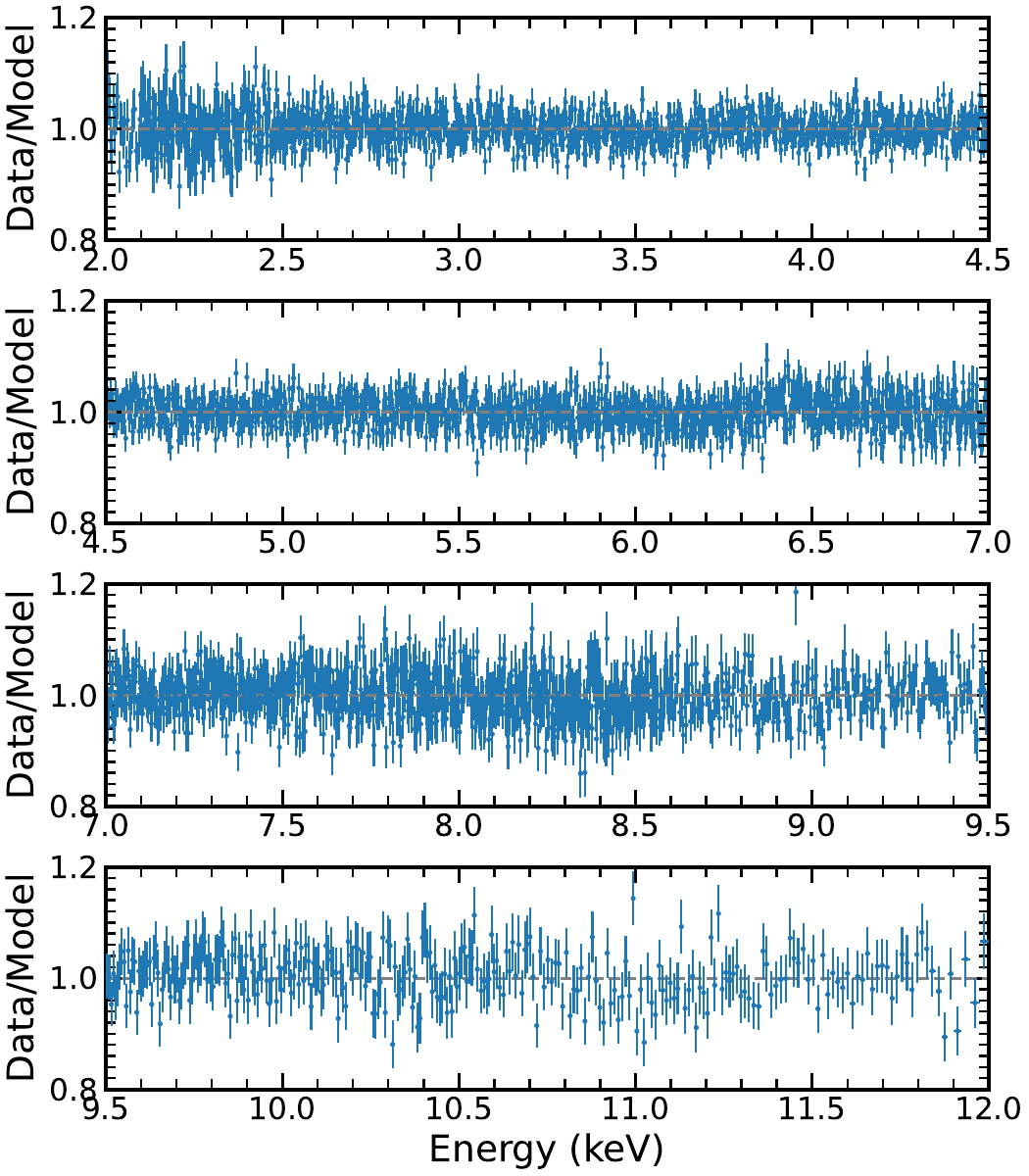}
    \caption{
        Data-to-model ratio plot for the Resolve-only time-averaged data fitted with Model 1. For visual clarity, the data are divided into four energy panels.
    }
    \label{rsl_resi}
\end{figure}

The best-fit parameters from the full dataset, the Resolve-only and the NuSTAR-only dataset are presented in Table~\ref{tab:bestfit_combined}. The residuals for the Resolve-only fit are shown in Figure~\ref{rsl_resi}. They indicate that the relativistic reflection model provides a good fit to the data. We also do not see any remaining narrow features, such as a narrow core around 6.4 keV that has been interpreted as distant reflection in some BH XRB systems.

A broken-power-law emissivity profile is required to fit the full dataset, with a steep inner emissivity index and a flat outer index. We find a rapidly spinning black hole ($a_*>0.98$) and a disk that extends down to the ISCO radius. The high black hole spin is consistent with previous studies using \nustar{} data from the 2015 outburst \citep{El-Batal2016ApJ...826L..12E, Xu2018ApJ...865..134X}.

The disk inclination angle is measured to be about $10^\circ$, lower than the value of about $50^\circ$ obtained from the simultaneous dataset. This low inclination angle measurement should be taken with caution due to the variability of the source across the \xrism{} observation. Using the 2015 \nustar{} data, \cite{El-Batal2016ApJ...826L..12E} constrained the inclination to be around $70^\circ$. Using the same data but a different reflection model, \cite{Liu2023ApJ...951..145L} obtained a value about $65^\circ$. Combining constraints from three \nustar{} observations in 2015, \cite{Draghis2024ApJ...969...40D} reported an inclination angle of $47_{-10}^{+11}{}^\circ$.

When fitting the full dataset with Model 1, including the \nustar{} data leads to an extreme value of the reflection fraction. Such a value is difficult to achieve for BH XRBs in the hard state. We find that this may be related to the high-energy tail. If we replace the \texttt{relxillCp} model with \texttt{relxill}, the inferred reflection fraction becomes slightly below 1 (see Table~\ref{tab:bestfit_combined}). The main difference between the two models is that the incident spectrum in \texttt{relxill} is described by a phenomenological power law with an exponential high-energy cutoff. The fit with the \texttt{relxill} model still returns a high BH spin parameter. However, the inferred disk inclination angle is moderately high ($\sim 60^\circ$).

\section{Discussion}
\label{discuss}

In this work, we present a new spectroscopic analysis of \xrism{} and \nustar{} observations of \src{} during its 2026 outburst. The high energy-resolution data from Resolve reveal a singly-peaked broad iron line profile without noticeable absorption or additional narrow emission line components. The \nustar{} data constrain the continuum and the Compton hump. The lack of absorption features in the X-ray band could be due to the fact that the source was in the hard state during these observations, where disk winds are typically weak or absent \citep{Ponti2012MNRAS.422L..11P, Parra2024A&A...681A..49P}. The other possibility is that the disk inclination angle is low and the equatorial disk winds do not intercept our line of sight.

Modeling the reflection features suggests a high black hole spin ($a_* > 0.98$ at the 90\% confidence level), consistent with previous studies using only \nustar{} data \citep{El-Batal2016ApJ...826L..12E, Xu2018ApJ...865..134X}. The high spin is inferred from fits to both the full dataset and the Resolve-only dataset. The combined constraint on the black hole spin reported by \cite{Draghis2024ApJ...969...40D} is $a_* = 0.849_{-0.221}^{+0.103}$ (1-$\sigma$ confidence level), which is slightly lower than our measurement but still consistent within the uncertainties. 

We also note that the spin measurement here considers only statistical uncertainties. Systematic uncertainties from the assumption in the coronal geometry and the reflection models are not included in the error bars \citep[e.g.,][]{Bambi2021}. A more detailed analysis of the data considering time-resolved spectroscopy and different reflection models will help to further investigate the systematic uncertainties in the spin measurement (Liu et al., in prep).

Across all tested datasets and models, the inner disk radius is constrained to be $<3\,R_{\rm g}$.
For the best fit to the simultaneous dataset, the unabsorbed X-ray flux in the 0.5--100~keV band is $1.66\times10^{-8}$~erg~cm$^{-2}$~s$^{-1}$, which is about twice the flux reported by \cite{El-Batal2016ApJ...826L..12E} during the 2015 outburst. Assuming a black hole mass of $10\,M_\odot$ and the distance measurement from Gaia, this corresponds to an Eddington ratio ($L_{\rm 0.5\text{--}100\,keV}/L_{\rm Edd}$) of 0.08\%--2.8\%. The large range reflects the substantial uncertainty in the distance measurement.
Assuming a distance of 25~kpc \citep{Casares2009ApJS..181..238C}, we obtain $L_{\rm 0.5\text{--}100,keV}/L_{\rm Edd} \sim 1$. Such a high Eddington ratio is unlikely for BH XRBs in the hard state.
From the hardness-intensity diagram (Figure~\ref{HID}), the source is likely in the bright hard state during the \xrism{} observation, before the possible hard-to-soft state transition. At this stage of the outburst, an inner disk radius close to the ISCO has also been suggested by other sources \citep{Garcia2015,Liu2023ApJ...950....5L,Fan2024ApJ...969...61F, Liu2026arXiv260300833L}, although there are still some remaining debates over these results \citep[e.g.,][]{Basak2016MNRAS.458.2199B, Zdziarski2022ApJ...928...11Z}.

We find that measurements of the disk inclination angle of \src{} from reflection modeling are model dependent. When fitting the simultaneous dataset, the inferred inclination ranges from about $30^\circ$ to $50^\circ$, depending on the reflection model used. Fitting the full dataset yields an inclination of either about $10^\circ$ or about $60^\circ$. We also fit the full dataset with Model 2 and obtain an inclination angle of $31.8^\circ \pm 0.5^\circ$, which is consistent with the measurement from the simultaneous dataset. However, this fit results in a \texttt{Cstat} value that is worse by 400 compared to Model 1. The inclination angle derived from the full dataset in this analysis should be treated with caution, as the source is variable during the \xrism{} observation.
In the literature, measurements of the disk inclination of \src{} also vary substantially, spanning roughly $30^\circ$ to $70^\circ$ \citep{El-Batal2016ApJ...826L..12E,Liu2023ApJ...951..145L,Draghis2024ApJ...969...40D}. This is not unique to \src{}. For example, the inferred disk inclination of XTE~J1752--223 can differ by up to $30^\circ$ depending on the reflection model used \citep{Garcia2018ApJ86425G, Connors2022ApJ...935..118C}. This uncertainty mainly arises from our limited knowledge of the coronal geometry and the insufficient signal-to-noise ratio to distinguish between different models. Given the high energy resolution and large number of counts in the XRISM data, a time-resolved analysis that fully exploits the available signal-to-noise ratio may be able to discriminate between models spectroscopically and provide a robust measurement of the disk inclination angle.

One other way to better constrain the disk inclination is to study the X-ray polarization properties of the source. The X-ray polarization signal from accreting BHs is sensitive to the disk inclination angle, with a higher inclination angle generally leading to a higher polarization degree \citep{Chandrasekhar1960ratr.book.....C, Schnittman2009ApJ...701.1175S, Schnittman2010ApJ...712..908S, Liu2026JHEAp..5100548L}. It could also provide independent BH spin measurements \citep{Dovvciak2008MNRAS.391...32D}. Such measurements have been made for a few sources using data from the Imaging X-ray Polarimetry Explorer \citep[IXPE,][]{Weisskopf2022JATIS...8b6002W}. The results are generally consistent with the reflection-based measurements \citep[e.g.,][]{Steiner2024ApJ...969L..30S, Marra2024A&A...684A..95M, Zhao2026ApJ...997L..12Z}. IXPE observed \src{} from 07 to 09 February 2026, two weeks after the \xrism{} observation reported in this work. At that time, the source was in a softer state with a photon index of $\sim 2.46$ and a QPO frequency of $\sim 5$ Hz \citep{Ravi:2026uan}. The polarization degree in the IXPE band (2–8 keV) is $\sim 4\%$, suggesting a moderate-to-high inclination angle \citep{Ravi:2026uan}. Constraining the inclination angle properly with the IXPE data requires more detailed modeling than using the overall polarization degree alone.

The iron abundance is found to be close to the solar value, consistent with previous studies \citep{El-Batal2016ApJ...826L..12E, Liu2023ApJ...951..145L}. We note that significantly super-solar iron abundances have been reported in many BH XRBs \citep{Parker2015ApJ...808....9P, Walton2017, Garcia2018ASPC, Buisson2019, Wang2020, Connors2022ApJ...935..118C, Liu2023ApJ...950....5L, Song2023MNRAS.526.6041S, Kumar2024MNRAS.532.2635K, Mall2024MNRAS.52712053M, Zdziarski2026ApJ...998L..37Z}. In some cases, allowing the disk density to exceed the canonical value of $10^{15}$~cm$^{-3}$ can recover an iron abundance close to solar \citep{Garcia2016, Tomsick2018, Jiang2019gx339, Chakraborty2021MNRAS.508..475C}. This is not the case for \src{}, as no super-solar iron abundance is required even when using traditional reflection models. In a \nustar{} sample of 36 BH systems, \cite{Draghis2025ApJ...989..227D} find iron abundances spanning from sub-solar to super-solar values. Although high-density reflection models can affect the inferred iron abundance, their impact on BH spin measurements is minor \citep{Draghis2025ApJ...989..227D}.

\begin{acknowledgements}
The authors thank the XRISM and NuSTAR teams for planning and conducting the observations. DJW acknowledges support from the Science and Technology Facilities Council (STFC; grant code ST/Y001060/1).
\end{acknowledgements}

\bibliography{ref}
\bibliographystyle{yahapj}



\end{document}